\documentclass[11pt,preprint,showpacs]{revtex4}
\usepackage{graphicx}
\usepackage{dcolumn}
\usepackage{bm}
\usepackage{amsmath}
\usepackage{amssymb}
\usepackage{amsfonts}

\begin{document}

\title{Momentum-space 3N Faddeev calculations of hadronic and
 electromagnetic reactions
 with proton-proton  Coulomb and three-nucleon forces included}

\author{H.~Wita{\l}a}
\affiliation{M. Smoluchowski Institute of Physics, Jagiellonian
University,
                    PL-30059 Krak\'ow, Poland}

\author{R.~Skibi\'nski}
\affiliation{M. Smoluchowski Institute of Physics, Jagiellonian
University,
                    PL-30059 Krak\'ow, Poland}

\author{J.~Golak}
\affiliation{M. Smoluchowski Institute of Physics, Jagiellonian
University,
                    PL-30059 Krak\'ow, Poland}

\author{W.\ Gl\"ockle}
\affiliation{Institut f\"ur theoretische Physik II,
Ruhr-Universit\"at Bochum, D-44780 Bochum, Germany}

\date{\today}

\begin{abstract}
We extend our approach to incorporate the proton-proton (pp)
Coulomb force into the three-nucleon (3N) momentum-space
Faddeev calculations of
elastic proton-deuteron (pd) scattering and breakup  to the case when
 also a three-nucleon force (3NF) is acting. In addition we formulate that
approach in the application to  electron- and $\gamma$-induced reactions on
$^3$He.
 The main new ingredient is a 3-dimensional screened pp  Coulomb
t-matrix obtained by  a numerical solution of a 3-dimensional
Lippmann-Schwinger  equation (LSE). The resulting
equations have the same structure as the Faddeev equations which
describe pd 
scattering without 3NF acting. That shows the practical feasibility of
both presented formulations.
\end{abstract}


\maketitle \setcounter{page}{1}

\section{Introduction}
 \label{intro}

The  long-range nature of the Coulomb force
prevents the application of the standard techniques developed for
short-range interactions   in  the analysis of nuclear
reactions involving two protons.
 One proposal to avoid the
difficulties  including the Coulomb force is to use a screened
Coulomb interaction and to reach the pure Coulomb limit through
application of a renormalisation procedure~\cite{Alt78,Alt96,Alt94,Alt2002}.

For  elastic pd
scattering first calculations, with modern nuclear forces and the
 exact Coulomb force in coordinate representation
included, have been achieved in a variational
hyperspherical harmonic approach~\cite{kievski}. Recently the
inclusion of the Coulomb force was undertaken also for the pd
breakup reaction using a screened pp Coulomb force
 in momentum space and in a partial wave basis~\cite{delt2005br}.
 To get the final predictions which can be compared to the
data, the limit to the unscreened situation has been performed
numerically applying a renormalization to the resulting
 3N on-shell amplitudes~\cite{delt2005el,delt2005br}.

One main concern in such
type of calculations is the application of a partial wave decomposition
to the long-ranged  Coulomb force. Even when screening is applied
it  seems reasonable to treat from the beginning the screened pp
Coulomb t-matrix without partial wave decomposition because the required
limit
 of vanishing screening leads necessarily to  a drastic increase
of the number of partial wave states involved. In  consequence
this  leads  to an explosion of the
number of 3N partial waves required for convergence.

Therefore  we developed in~\cite{elascoul,brcoul} a novel approach
to include the pp Coulomb
force into the momentum space 3N Faddeev calculations.
 It is based on a standard
 formulation for short range forces and relies on the screening  of the
long-range Coulomb interaction. In order to avoid  all uncertainties
connected with the application of the partial wave expansion,
 inadequate  when working
with long-range forces, we used directly the 3-dimensional pp screened
Coulomb t-matrix.
We demonstrated in \cite{elascoul,brcoul}  the
feasibility of that approach in the case of elastic pd scattering and
breakup
using a simple dynamical model for the nuclear part of the interaction.

In the present paper we extend that  approach to include a 3NF into
that formulation.
Also we show how that formulation can be applied to
electromagnetic processes induced by electrons or $\gamma$'s on $^3$He.

In section \ref{form} for the convenience of the reader  we shortly
 describe  the main points of the formalism outlined in detail
in \cite{elascoul,brcoul} for the case of 3N Faddeev calculations with
pairwise forces only and extend the corresponding equations to the case
when a 3NF is also acting.
 In section \ref{electro} we apply
 that formulation to electromagnetic reactions on $^3$He.
 The summary is given in section
\ref{sumary}.

\section{Faddeev equations with screened pp Coulomb force}
\label{form}

When only pairwise forces are acting we use the Faddeev equation
in the form~\cite{physrep96,gloeckle83}
\begin{eqnarray}
T| \Phi> =  t P | \Phi> +  t P G_0 T| \Phi> ~,
\label{13a}
\end{eqnarray}
where the permutation operator $ P $ is defined in terms of transposition
operators $P_{ij}$ of nucleons $i$ and $j$,
 $ P =P_{12} P_{23} + P_{13} P_{23} $, $G_0$ is the free 3N propagator,
 and $ |\Phi>$ is the initial state composed of a deuteron state and a momentum
 eigenstate of the proton.
Knowing $T| \Phi>$ the breakup as well as the elastic pd scattering amplitudes
 can be gained in the standard manner~\cite{physrep96}. The physical
 content of Eq.~(\ref{13a}) is revealed after iterating it. The
 resulting multiple-scattering series contains all possible
 rescattering contributions induced by interactions of three nucleons
 and free propagation in between.

We use our standard momentum space partial wave basis $|pq\tilde {\alpha}>$
\begin{eqnarray}
|p q \tilde {\alpha}> \equiv   |pq(ls)j(\lambda
\frac {1} {2})I (jI)J (t \frac {1} {2})T>\label{51}
\end{eqnarray}
and  distinguish between the partial wave states $|pq\alpha>$ and
$|pq\beta>$. The $|pq\alpha>$ are states with
total 2N angular momentum $j$ below some value $j_{max}$: $j \le
j_{max}$,
 in which the nuclear, $V_N$, as well as the pp screened Coulomb
interaction, $V_c^R$
 (in isospin $t=1$ states only), are acting. In the
states $|pq\beta>$ with $j > j_{max}$ only $V_c^R$ is
acting in the pp subsystem. The states $|pq\alpha>$ and
$|pq\beta>$ form a complete set of states
\begin{equation}
\int {p^2 dpq^2 dq} \sum\limits_{\tilde \alpha}  {\left| {pq{\tilde \alpha} }
\right\rangle } \left\langle {pq{\tilde \alpha} } \right| =
\int {p^2 dpq^2 dq} (\sum\limits_\alpha  {\left| {pq\alpha }
\right\rangle } \left\langle {pq\alpha } \right| +
\sum\limits_\beta  {\left| {pq\beta } \right\rangle } \left\langle
{pq\beta } \right|) = {\rm I} \\~ .
\label{52}
\end{equation}
Projecting Eq.(\ref{13a}) for $ T| \Phi >$ on the $|pq\alpha>$ and
$|pq\beta>$ states one gets the following system of coupled
integral equations
\begin{eqnarray}
 \left\langle {pq\alpha } \right|T\left| {\Phi } \right\rangle  &=&
 \left\langle {pq\alpha } \right|t_{N + c}^R P\left| {\Phi } \right\rangle
\cr
 &+& \left\langle {pq\alpha } \right|t_{N + c}^R PG_0 \sum\limits_{\alpha '}
 {\int {p'^2 dp'q'^2 dq'\left| {p'q'\alpha '} \right\rangle \left\langle
 {p'q'\alpha '} \right|} } T\left| {\Phi } \right\rangle \cr
 &+& \left\langle {pq\alpha } \right|t_{N + c}^R PG_0 \sum\limits_{\beta '}
 {\int {p'^2 dp'q'^2 dq'\left| {p'q'\beta '} \right\rangle \left\langle
 {p'q'\beta '} \right|} } T\left| {\Phi } \right\rangle ~,
\label{53} \\
 \left\langle {pq\beta } \right|T\left| {\Phi } \right\rangle  &=&
 \left\langle {pq\beta } \right|t_c^R P\left| {\Phi } \right\rangle \cr
  &+& \left\langle {pq\beta } \right|t_c^R PG_0 \sum\limits_{\alpha '}
  {\int {p'^2 dp'q'^2 dq'\left| {p'q'\alpha '} \right\rangle
  \left\langle {p'q'\alpha '} \right|} } T\left| {\Phi } \right\rangle \cr
  &+& \left\langle {pq\beta } \right|t_c^R PG_0 \sum\limits_{\beta '}
  {\int {p'^2 dp'q'^2 dq'\left| {p'q'\beta '} \right\rangle
  \left\langle {p'q'\beta '} \right|} } T\left| {\Phi } \right\rangle ~,
 \label{54}
\end{eqnarray}
where $t_{N+c}^R$ and $t_c^R$ are t-matrices generated through a LSE by
the interactions $V_N+V_c^R$ and $V_c^R$, respectively.
 Namely for states $|\alpha >$ with two-nucleon subsystem total isospin
$t=1$ the corresponding t-matrix element
$<p\alpha |t_{N+c}^R(E-\frac {3} {4m}q^2)|p'\alpha'>$ is a linear
combination of the pp, $t_{pp+c}^R$, and the neutron-proton (np),
$t_{np}$, $t=1$ t-matrices, which are generated by the interactions
$V_{pp}^{strong}+V_c^R$ and $V_{np}^{strong}$,
respectively. The
coefficients of that combination depend on the total 3N isospin $T$ and
$T'$
of the states $|\alpha >$ and $|\alpha' >$~\cite{elascoul,wit91}:
\begin{eqnarray}
<t=1T=\frac{1} {2} |t_{N+c}^R|t'=1T'=\frac{1}
{2}>
& =&  \frac {1} {3} t_{np} + \frac {2} {3} t_{pp+c}^{R} \cr
<t=1T=\frac{3} {2} |t_{N+c}^R|t'=1T'=\frac{3}
{2}>
& =&  \frac {2} {3} t_{np} + \frac {1} {3} t_{pp+c}^{R} \cr
<t=1T=\frac{1} {2} |t_{N+c}^R|t'=1T'=\frac{3}
{2}>
& =&  \frac {\sqrt{2}} {3}( t_{np} - t_{pp+c}^{R}) \cr
<t=1T=\frac{3} {2} |t_{N+c}^R|t'=1T'=\frac{1}
{2}>
& =&  \frac {\sqrt{2}} {3}( t_{np} - t_{pp+c}^{R})    ~.
\label{8}
\end{eqnarray}
For the isospin $t=0$, in which case
$T=T'=\frac {1} {2}$:
\begin{eqnarray}
<t=0T=\frac{1} {2} |t_{N+c}^R|t'=0T'=\frac{1}
{2}>
& =& t_{np} ~.
\label{8a}
\end{eqnarray}
In the case of $t_c^R$ only the screened pp Coulomb force $V_c^R$ is acting.

The third term  on the right hand side of (\ref{54}) is
proportional to $<pq\beta|t_c^RPG_0|p'q'\beta'><p'q'\beta'|t_c^ R$. A direct
calculation of its isospin part shows that independently from the
value of the total isospin $T$ it vanishes~\cite{elascoul}.

Inserting $<pq\beta|T|\Phi>$ from (\ref{54}) into (\ref{53}) one
gets
\begin{eqnarray}
 \left\langle {pq\alpha } \right|T\left| {\Phi } \right\rangle  &=&
 \left\langle {pq\alpha } \right|t_{N + c}^R P\left| {\Phi } \right\rangle
 + \left\langle {pq\alpha } \right|t_{N + c}^R PG_0 t_c^R P\left| {\Phi }
\right\rangle \cr
  &-& \left\langle {pq\alpha } \right|t_{N + c}^R PG_0 \sum\limits_{\alpha
'}
  {\int {p'^2 dp'q'^2 dq'\left| {p'q'\alpha '} \right\rangle
  \left\langle {p'q'\alpha '} \right|} } t_c^R P\left| {\Phi } \right\rangle
\cr
  &+& \left\langle {pq\alpha } \right|t_{N + c}^R PG_0 \sum\limits_{\alpha
'}
  {\int {p'^2 dp'q'^2 dq'\left| {p'q'\alpha '} \right\rangle
  \left\langle {p'q'\alpha '} \right|} } T\left| {\Phi }
  \right\rangle \cr
  &+& \left\langle {pq\alpha } \right|t_{N + c}^R PG_0 t_c^R PG_0
\sum\limits_{\alpha '}
  {\int {p'^2 dp'q'^2 dq'\left| {p'q'\alpha '} \right\rangle
  \left\langle {p'q'\alpha '} \right|} } T\left| {\Phi } \right\rangle \cr
  &-& \left\langle {pq\alpha } \right|t_{N + c}^R PG_0  \sum\limits_{\alpha
'}
  {\int {p'^2 dp'q'^2 dq'\left| {p'q'\alpha '} \right\rangle
  \left\langle {p'q'\alpha '} \right|} } t_c^R PG_0 \cr
  && \sum\limits_{\alpha'' }
  {\int {p''^2 dp''q''^2 dq''\left| {p''q''\alpha'' } \right\rangle
  \left\langle {p''q''\alpha'' } \right|} } T\left| {\Phi } \right\rangle ~.
  \label{55}
 \end{eqnarray}
This is a coupled set of integral equations in the space of  the states
$|\alpha>$ only, which incorporates the contributions of
the pp Coulomb interaction from all partial wave states up to
infinity.
 It can be solved by iteration and Pade
summation~\cite{physrep96,elascoul}.

When compared to our standard treatment without screened Coulomb
force~\cite{physrep96}  there are two new leading terms:
$<pq\alpha|t_{N+c}^RPG_0t_c^RP|\Phi>$  and
-$<pq\alpha|t_{N+c}^RPG_0|\alpha'><\alpha'|t_c^RP|\Phi>$. The
first term must be calculated using directly
the $3$-dimensional screened Coulomb t-matrix $t_c^R$, while the
second term requires only the partial wave projected screened Coulomb
t-matrix elements in the  $|\alpha>$ channels. The kernel also contains
two new terms: the term
$<pq\alpha|t_{N+c}^RPG_0t_c^RPG_0|\alpha'><\alpha'|T|\Phi>$ must again be
calculated with a 3-dimensional screened Coulomb t-matrix $t_c^R$,
while the second
one,
-$<pq\alpha|t_{N+c}^RPG_0|
\alpha'><\alpha'|t_c^RPG_0|\alpha''><\alpha''|T|\Phi>$,  involves only
 the partial wave projected screened Coulomb t-matrix elements in the
$|\alpha>$ channels. The calculation of those new terms with the
partial wave projected Coulomb t-matrices follows our standard
procedure~\cite{physrep96}. Namely the  two sub-kernels $t_{N+c}^RPG_0$ and
$t_c^RPG_0$ are applied consecutively  on the corresponding state.
The detailed expressions how to calculate the new terms with the  3-dimensional
screened Coulomb t-matrix  are given in Appendix A of Ref.~\cite{elascoul}.

The transition amplitude for
 breakup,  $<\Phi_0|U_0|\Phi>$,  is given
in terms of $T\left| {\Phi } \right\rangle$ by~\cite{gloeckle83,physrep96}
\begin{eqnarray}
 \left\langle {\Phi _0 } \right|U_0 \left| {\Phi } \right\rangle  &=&
 \left\langle {\Phi _0 } \right|(1 + P)T\left| {\Phi } \right\rangle
\label{56}
 \end{eqnarray}
where $ | \Phi_0> \equiv | \vec p
\vec q m_1 m_2 m_3 \nu_1 \nu_2 \nu_3 > $ is the free state and the Jacobi
momenta $\vec p$ and $\vec q$ specify completely a particular exclusive breakup
configuration of three outgoing nucleons.
 The permutations acting in momentum-, spin-, and isospin-spaces
can be applied to the bra-state $< \Phi_0| = < \vec p
\vec q m_1 m_2 m_3 \nu_1 \nu_2 \nu_3 | $ changing the sequence of
nucleons spin and isospin magnetic
quantum numbers $ m_i$ and  $\nu_i$ and leading to well known linear
combinations of the Jacobi momenta $\vec p$ and $\vec q$. Thus
evaluating (\ref{56}) it is sufficient to regard the general
amplitudes $< \vec p \vec q m_1 m_2 m_3 \nu_1 \nu_2 \nu_3 |
T\left| {\Phi }\right\rangle \equiv \left\langle {\vec p\vec q~}
\right|T\left| {\Phi }
\right\rangle $. Using Eq.~(\ref{54}) and the completeness relation (\ref{52})
one gets:
\begin{eqnarray}
&&\left\langle {\vec p\vec q~} \right|T\left| {\Phi }
\right\rangle  = \left\langle {\vec p\vec q~}
\right|\sum\limits_{\alpha '} {\int {p'^2 dp'q'^2 dq'\left|
{p'q'\alpha '} \right\rangle \left\langle {p'q'\alpha '} \right|}
} T\left| {\Phi } \right\rangle \cr && - \left\langle {\vec
p\vec q~} \right|\sum\limits_{\alpha '} {\int {p'^2 dp'q'^2
dq'\left| {p'q'\alpha '} \right\rangle \left\langle {p'q'\alpha '}
\right|} } t_c^R P\left| {\Phi } \right\rangle \cr &&-
\left\langle {\vec p\vec q~} \right|\sum\limits_{\alpha '} {\int
{p'^2 dp'q'^2 dq'\left| {p'q'\alpha '} \right\rangle \left\langle
{p'q'\alpha '} \right|} } t_c^R PG_0
 \sum\limits_{\alpha
''} {\int {p''^2 dp''q''^2 dq''\left| {p''q''\alpha ''}
\right\rangle \left\langle {p''q''\alpha ''} \right|} } T\left|
{\Phi } \right\rangle \cr && + \left\langle {\vec p\vec q~}
\right|t_c^R P\left| {\Phi } \right\rangle  + \left\langle
{\vec p\vec q~} \right|t_c^R PG_0 \sum\limits_{\alpha '} {\int
{p'^2 dp'q'^2 dq'\left| {p'q'\alpha '} \right\rangle \left\langle
{p'q'\alpha '} \right|} } T\left| {\Phi } \right\rangle ~.
\label{69}
\end{eqnarray}

It follows, that in addition to the amplitudes $<pq\alpha|T|\Phi >$ also the
partial wave projected amplitudes $<pq\alpha|t_c^RP|\Phi >$ and
$<pq\alpha|t_c^RPG_0|\alpha'><\alpha'|T|\Phi>$ are required. The
expressions for the contributions of these three terms to the
transition amplitude for the breakup reaction are
given in Appendix B of Ref.~\cite{elascoul}.

The last two terms in (\ref{69}) again must be calculated using
directly the $3$-dimensional screened Coulomb t-matrices. In
Appendix  C 
of Ref.~\cite{elascoul} the expression for
$\left\langle {\vec p\vec q~} \right|t_c^R
P\left| {\Phi } \right\rangle$ is given 
 and in Appendix D  for
 the last matrix element $<\vec p \vec q~ |t_c^RPG_0|\alpha
'><\alpha '|T|\Phi>$.

The transition amplitude for elastic scattering, U, contains in addition
to all rescatterings $PT$ also a direct exchange term  $PG_0^{-1}$
and is given by
\begin{eqnarray}
 \left\langle {\Phi' } \right|U \left| {\Phi } \right\rangle  &=&
 \left\langle {\Phi'} \right|PG_0^{-1} + PT\left| {\Phi }
 \right\rangle ~,
\label{56_elas}
 \end{eqnarray}
where the  outgoing proton-deuteron state $\left| {\Phi' } \right\rangle$
differs from $\left| {\Phi } \right\rangle$ by the direction of the
relative proton-deuteron momentum. It can be obtained by quadrature
using (\ref{69}).

It was shown in \cite{elascoul} that the elastic pd scattering
amplitude has a well-defined screening limit and does not require
renormalisation. To get the physical breakup amplitude,
however,  it is
unavoidable to perform the renormalisation of the pp half-shell
t-matrices \cite{brcoul}.

When on top of pairwise forces three nucleons interact also
through a 3NF additional rescatterings, generated by that 3NF, appear
in the  multiple-scattering series and the Faddeev equation for
 the state $T \left| {\Phi} \right\rangle$ changes to \cite{huber}:
\begin{eqnarray}
T \left| {\Phi} \right\rangle = tP\left| \Phi  \right\rangle
+ (1 + tG_0 )V_4^{(1)} (1 + P)\left| \Phi  \right\rangle  +
tPG_0 T\left| {\Phi} \right\rangle
+ (1 + tG_0 )V_4^{(1)} (1 + P)G_0 T\left| {\Phi} \right\rangle ~.
\label{e3nf.1}
\end{eqnarray}
The 3N force $V_4$ is naturaly split into 3 parts
\begin{eqnarray}
V_4 = V_4^{(1)} +  V_4^{(2)} +  V_4^{(3)}
\label{e3nf_split}
\end{eqnarray}
where  $V_4^{(i)}$ is  symmetrical under the exchange of nucleons j and k
  (i,j,k=1, 2, 3, $i \ne j \ne k$). Such a splitting is always possible
  and in  case of the $\pi-\pi$ exchange 3NF
  corresponds to the  three   possible choices of the nucleon undergoing
  off-shell $\pi$N scattering.
 Eq.~(\ref{e3nf.1}) contains two new terms: one leading term and one
 in the kernel. They reflect additional contributions
to the multiple-scattering series caused
by the  3NF.

Performing analogous steps as for (\ref{13a}) and starting with
the projection of  ($\ref{e3nf.1}$)  on the $| pq\alpha \rangle$ and
$| pq\beta \rangle$ states one gets:
\begin{eqnarray}
&&  \left\langle {{pq\alpha }}
 \mathrel{\left | {\vphantom {{pq\alpha } T}}
 \right. \kern-\nulldelimiterspace}
 {T} |\Phi \right\rangle  = \left\langle {pq\alpha } \right|t_{N+c}^R P\left|
\Phi  \right\rangle  + \left\langle {pq\alpha } \right|(1 + t_{N+c}^R G_0 )
V_4^{(1)} (1 + P)\left| \Phi  \right\rangle  \hfill \cr &&
   + \left\langle {pq\alpha } \right|t_{N+c}^R PG_0 \left| \alpha'
\right\rangle \left\langle {\alpha' }
 \mathrel{\left | {\vphantom {\alpha'  T}}
 \right. \kern-\nulldelimiterspace}
 {T} |\Phi \right\rangle  + \left\langle {pq\alpha }
\right|t_{N+c}^R PG_0 \left| \beta'  \right\rangle \left\langle {\beta' }
 \mathrel{\left | {\vphantom {\beta'  T}}
 \right. \kern-\nulldelimiterspace}
 {T} |\Phi \right\rangle  \hfill \cr &&
   + \left\langle {pq\alpha } \right|(1 + t_{N+c}^R G_0 )V_4^{(1)}
(1 + P)G_0 \left| \alpha'  \right\rangle \left\langle {\alpha' }
 \mathrel{\left | {\vphantom {\alpha'  T}}
 \right. \kern-\nulldelimiterspace}
 {T} |\Phi \right\rangle  \cr &&
+ \left\langle {pq\alpha }
\right|(1 + t_{N+c}^R G_0 )V_4^{(1)} (1 + P)G_0 \left| \beta'
\right\rangle \left\langle {\beta' }
 \mathrel{\left | {\vphantom {\beta'  T}}
 \right. \kern-\nulldelimiterspace}
 {T} |\Phi \right\rangle  \hfill  ~,
 \label{e3nf.2}
\end{eqnarray}
and
\begin{eqnarray}
&&  \left\langle {{pq\beta }}
 \mathrel{\left | {\vphantom {{pq\beta } T}}
 \right. \kern-\nulldelimiterspace}
 {T} |\Phi \right\rangle  = \left\langle {pq\beta } \right|t_c^R P\left|
\Phi  \right\rangle  + \left\langle {pq\beta } \right|(1 + t_c^R G_0 )
V_4^{(1)} (1 + P)\left| \Phi  \right\rangle  \hfill \cr &&
   + \left\langle {pq\beta } \right|t_c^R PG_0 \left| \alpha'
\right\rangle \left\langle {\alpha' }
 \mathrel{\left | {\vphantom {\alpha'  T}}
 \right. \kern-\nulldelimiterspace}
 {T} |\Phi \right\rangle  + \left\langle {pq\beta } \right|t_c^R PG_0
\left| \beta'  \right\rangle \left\langle {\beta' }
 \mathrel{\left | {\vphantom {\beta'  T}}
 \right. \kern-\nulldelimiterspace}
 {T} |\Phi \right\rangle  \hfill \cr &&
   + \left\langle {pq\beta } \right|(1 + t_c^R G_0 )V_4^{(1)}
(1 + P)G_0 \left| \alpha'  \right\rangle \left\langle {\alpha' }
 \mathrel{\left | {\vphantom {\alpha'  T}}
 \right. \kern-\nulldelimiterspace}
 {T} |\Phi \right\rangle  \cr&&
+ \left\langle {pq\beta } \right|
(1 + t_c^R G_0 )V_4^{(1)} (1 + P)G_0 \left| \beta'  \right\rangle
\left\langle {\beta' }
 \mathrel{\left | {\vphantom {\beta'  T}}
 \right. \kern-\nulldelimiterspace}
 {T} |\Phi \right\rangle  \hfill ~.
 \label{e3nf.3}
\end{eqnarray}
Here and in the following we shortened our notation by neglecting the
summation sign over intermediate states $|\alpha'>$ ($|\beta'>$)
  and the integration sign over the  corresponding
Jacobi momenta $p'$ and $q'$.
  Therefore whenever a projection operator
  $|\alpha'><\alpha'|$ appears in intermediate states   means
  that the following summation and integrations must be performed:
\begin{eqnarray}
|\alpha'><\alpha'| &\equiv&  \sum\limits_{\alpha '} {\int
{p'^2 dp'q'^2 dq'\left| {p'q'\alpha '} \right\rangle \left\langle
{p'q'\alpha '} \right|} } ~.
\end{eqnarray}

Since a 3NF is short-ranged its matrix elements containing $| \beta >$
channels vanish:
\begin{eqnarray}
\langle \alpha  | V_4^{(1)} (1 + P) | \beta \rangle  =
\langle \beta |V_4^{(1)} (1 + P)| \alpha \rangle  =\langle \beta  |
V_4^{(1)} (1 + P) | \beta \rangle  = 0 ~.
\label{e.3nf_zero}
\end{eqnarray}
Thus in the $|\beta >$ channels only the pp Coulomb force is acting and
therefore (\ref{e3nf.3}) reduces to:
\begin{eqnarray}
&&  \langle pq\beta
 | T |\Phi \rangle  = \langle pq\beta  | t_c^RP | \Phi  \rangle
+ \langle pq\beta
 |t_c^R PG_0 | \alpha'  \rangle \langle \alpha'
 | T | \Phi\rangle  + \langle pq\beta  |t_c^R PG_0
 | \beta'  \rangle \langle \beta' | T | \Phi \rangle   \cr &&
   = \langle pq\beta  | t_c^R P | \Phi  \rangle  + \langle pq\beta | t_c^R PG_0
 | \alpha'  \rangle \langle \alpha' | T | \Phi \rangle  ~.
 \label{e3nf.4}
\end{eqnarray}
Again we have used the fact that the third term in (\ref{e3nf.4}) vanishes
(see also remark after (\ref{8a})).
 Also note that $ t_c^R $ is diagonal in the high partial waves and 
consequently $ V_4^{(1)}$ does not contribute.

Inserting (\ref{e3nf.4}) into  (\ref{e3nf.2}) one gets:
\begin{eqnarray}
&&  \left\langle {{pq\alpha }}
| {T} | \Phi\right\rangle  = \left\langle {pq\alpha } \right|t_{N +
  c}^R P\left|
\Phi  \right\rangle  + \left\langle {pq\alpha } \right|
(1 + t_{N + c}^R G_0 )V_4^{(1)} (1 + P)\left| \Phi  \right\rangle  \hfill \cr &&
   + \left\langle {pq\alpha } \right|t_{N + c}^R PG_0 \left| \alpha'
\right\rangle \left\langle {\alpha' }
 \mathrel{\left | {\vphantom {\alpha'  T}}
 \right. \kern-\nulldelimiterspace}
  {T} | \Phi \right\rangle  + \left\langle {pq\alpha } \right|t_{N +
    c}^R PG_0
\left| \beta'  \right\rangle [\left\langle {\beta' } \right|t_c^R P\left|
\Phi  \right\rangle  + \left\langle {\beta' } \right|t_c^R PG_0 \left|
\alpha'  \right\rangle \left\langle {\alpha' }
 \mathrel{\left | {\vphantom {\alpha'  T}}
 \right. \kern-\nulldelimiterspace}
  {T} | \Phi \right\rangle ] \hfill \cr &&
   + \left\langle {pq\alpha } \right|(1 + t_{N + c}^R G_0 )V_4^{(1)}
(1 + P)G_0 \left| \alpha'  \right\rangle \left\langle {\alpha' }
 \mathrel{\left | {\vphantom {\alpha'  T}}
 \right. \kern-\nulldelimiterspace}
  {T} | \Phi \right\rangle  \hfill \cr &&
   + \left\langle {pq\alpha } \right|(1 + t_{N + c}^R G_0 )V_4^{(1)}
(1 + P)G_0 \left| \beta'  \right\rangle [\left\langle {\beta' }
\right|t_c^R P\left| \Phi  \right\rangle  + \left\langle {\beta' }
\right|t_c^R PG_0 \left| \alpha'  \right\rangle \left\langle {\alpha' }
 \mathrel{\left | {\vphantom {\alpha'  T}}
 \right. \kern-\nulldelimiterspace}
  {T} | \Phi \right\rangle ] \hfill  ~.
  \label{e3nf.5}
\end{eqnarray}
Thus
\begin{eqnarray}
&&  \left\langle {{pq\alpha }}
 \mathrel{\left | {\vphantom {{pq\alpha } T}}
 \right. \kern-\nulldelimiterspace}
  {T} | \Phi  \right\rangle  = \left\langle {pq\alpha } \right|t_{N +
    c}^R P\left|
\Phi  \right\rangle  + \left\langle {pq\alpha } \right|(1 + t_{N +c}^R G_0 )
V_4^{(1)} (1 + P)\left| \Phi  \right\rangle  \hfill \cr &&
   + \left\langle {pq\alpha } \right|t_{N + c}^R PG_0 \left| \beta'
\right\rangle \left\langle \beta'  \right|t_c^R P\left| \Phi  \right\rangle
+ \left\langle {pq\alpha } \right|(1 + t_{N + c}^R G_0 )V_4^{(1)} (1 + P)G_0
\left|
\beta'  \right\rangle \left\langle \beta'  \right|t_c^R P\left| \Phi
\right\rangle  \hfill \cr &&
   + \left\langle {pq\alpha } \right|t_{N + c}^R PG_0 \left| \alpha'
\right\rangle \left\langle {\alpha' }
 \mathrel{\left | {\vphantom {\alpha'  T}}
 \right. \kern-\nulldelimiterspace}
  {T} | \Phi \right\rangle  + \left\langle {pq\alpha } \right|t_{N +c}^R PG_0
 \left|
 \beta'  \right\rangle \left\langle \beta'  \right|t_c^R PG_0 \left| \alpha''
\right\rangle \left\langle {\alpha'' }
 \mathrel{\left | {\vphantom {\alpha''  T}}
 \right. \kern-\nulldelimiterspace}
  {T} | \Phi  \right\rangle  \hfill \cr &&
   + \left\langle {pq\alpha } \right|(1 + t_{N + c}^R G_0 )V_4^{(1)} (1 + P)G_0
\left| \alpha'  \right\rangle \left\langle {\alpha' }
 \mathrel{\left | {\vphantom {\alpha'  T}}
 \right. \kern-\nulldelimiterspace}
  {T} | \Phi \right\rangle  \hfill \cr &&
   + \left\langle {pq\alpha } \right|(1 + t_{N + c}^R G_0 )V_4^{(1)} (1 + P)G_0
\left| \beta'  \right\rangle \left\langle \beta'  \right|t_c^R PG_0 \left|
\alpha'  \right\rangle \left\langle {\alpha' }
 \mathrel{\left | {\vphantom {\alpha'  T}}
 \right. \kern-\nulldelimiterspace}
  {T} | \Phi \right\rangle  \hfill  ~.
   \label{e3nf.6}
\end{eqnarray}
Due to (\ref{e.3nf_zero}) the second term in the second line of
(\ref{e3nf.6}) and the last term
of (\ref{e3nf.6}) can be dropped. Using
the completeness relation (\ref{52}) for the $| \alpha >$ and $| \beta >$
states one finally  is
left with the  coupled set of integral equations in the space of $|\alpha >$
 channels only:
\begin{eqnarray}
&&  \left\langle {{pq\alpha }}
 \mathrel{\left | {\vphantom {{pq\alpha } T}}
 \right. \kern-\nulldelimiterspace}
 {T} | \Phi \right\rangle  = \left\langle {pq\alpha } \right|t_{N +c}^R P\left|
\Phi  \right\rangle  + \left\langle {pq\alpha } \right|(1 + t_{N +c}^R G_0 )
V_4^{(1)} (1 + P)\left| \Phi  \right\rangle  \hfill \cr &&
   + \left\langle {pq\alpha } \right|t_{N + c}^R PG_0 t_c^R P\left| \Phi
\right\rangle  - \left\langle {pq\alpha } \right|t_{N + c}^R PG_0 \left|
\alpha'  \right\rangle \left\langle \alpha'  \right|t_c^R P\left| \Phi
\right\rangle  \hfill \cr &&
   + \left\langle {pq\alpha } \right| t_{N + c}^R PG_0 \left|
\alpha'  \right\rangle \left\langle {\alpha' }
 \mathrel{\left | {\vphantom {\alpha'  T}}
 \right. \kern-\nulldelimiterspace}
  {T}| \Phi \right\rangle  \hfill \cr &&
   + \left\langle {pq\alpha } \right|t_{N + c}^R PG_0 t_c^R PG_0 \left| \alpha'
\right\rangle \left\langle {\alpha' }
 \mathrel{\left | {\vphantom {\alpha'  T}}
 \right. \kern-\nulldelimiterspace}
  {T} | \Phi \right\rangle  - \left\langle {pq\alpha } \right|t_{N +c}^R PG_0
\left| \alpha'  \right\rangle \left\langle \alpha'
\right|t_c^R PG_0 \left| \alpha''  \right\rangle \left\langle {\alpha'' }
 \mathrel{\left | {\vphantom {\alpha''  T}}
 \right. \kern-\nulldelimiterspace}
  {T}| \Phi \right\rangle  \hfill \cr &&
   + \left\langle {pq\alpha } \right|(1 + t_{N + c}^R G_0 )V_4^{(1)}
(1 + P)G_0 \left| \alpha'  \right\rangle \left\langle {\alpha' }
 \mathrel{\left | {\vphantom {\alpha'  T}}
 \right. \kern-\nulldelimiterspace}
  {T} | \Phi \right\rangle  \hfill  ~.
  \label{e3nf.9}
\end{eqnarray}
Comparing it to  equation (\ref{55}) with 2-body forces
only,  there is one additional contribution in the leading
term, $\left\langle {pq\alpha } \right|(1 + t_{N + c}^R G_0 )
V_4^{(1)} (1 + P)\left| \Phi  \right\rangle $,
and one in the  kernel,
$\left\langle {pq\alpha } \right|(1 + t_{N + c}^R G_0 )V_4^{(1)}
(1 + P)G_0 \left| \alpha'  \right\rangle \left\langle {\alpha' }
 | {T} | \Phi \right\rangle $,
both containing   $V_4^{(1)} (1 + P)$.

The transition amplitude for breakup is again given by (\ref{56})
 and for elastic scattering two new terms driven by
 $V_4^{(1)} (1 + P)$ appear \cite{huber}:
\begin{eqnarray}
 \left\langle {\Phi' } \right| U \left| {\Phi } \right\rangle
=  \left\langle {\Phi' } \right| PG_0^{ - 1}  + PT + V_4^{(1)} (1 + P)
  + V_4^{(1)} (1 + P)G_0 T \left| {\Phi } \right\rangle ~.
\end{eqnarray}

Since the structure of the set (\ref{e3nf.9}) is analogous to the
structure of the set
(\ref{55}) describing pd scattering when only pairwise forces are
acting, it follows that the inclusion of the 3NF 
 into the Faddeev calculations of
pd elastic scattering and breakup reaction requires no new matrix elements
and numerical tools beyond those used in \cite{elascoul,brcoul} and
\cite{huber}.

\section{The electromagnetic reactions on $^3$He}
\label{electro}

It was shown in \cite{gol_rep} that the basic equations describing
reactions on $^3$He induced by photons or electrons
 have the same structure as the 3N continuum Faddeev equations (\ref{13a})
 and (\ref{e3nf.1}). The new physical ingredient is
 the photon absorption operator, lets call it  {\bf O} \cite{gol_rep}.
 For  a complete breakup of $^3$He induced by  photons the nuclear matrix
element N, from which all observables can be determined, is given by
an auxiliary state $|U >$
which fulfills Faddeev type equation:
\begin{eqnarray}
&&  \left| U \right\rangle  = [tG_0  + \frac{1}
{2}(P + 1)V_4^{(1)} G_0 (1 + tG_0 )](1 + P) {\bf O} \left| \Psi_i
\right\rangle  \hfill \cr &&
   + (tG_0 P + \frac{1}
{2}(P + 1)V_4^{(1)} G_0 (1 + tG_0 )P)\left|  U
\right\rangle  \hfill
\label{e_el.2a}
\end{eqnarray}
Then 
\begin{eqnarray}
N = \left\langle {\Phi _0 } \right|(1 + tG_0 )(1 + P) {\bf O} \left| \Psi_i
\right\rangle  + \left\langle {\Phi _0 } \right|(1 + tG_0 )P
\left| U \right\rangle ~.
\label{e_el.3}
\end{eqnarray}
Here $| \Psi_i >$ is the initial $^3$He bound state and $|\Phi _0>$
  is the fully antisymmetrized free state of three outgoing nucleons,
  given in terms of their Jacobi momenta and spin and isospin quantum
  numbers.

For the pd breakup of $^3$He the nuclear matrix element is 
given \cite{gol_rep}  by
\begin{eqnarray}
N_{pd} = \left\langle {\Phi _q } \right|(1 + P) {\bf O} \left| \Psi_i
\right\rangle  + \left\langle {\Phi _q } \right|P
\left|  U \right\rangle
\label{e_el.4}
 \end{eqnarray}
where the final state is determined by the proton-deuteron relative
momentum eigenstate $ |\vec q >$   and the deuteron wave function
$|\phi_d >$:
\begin{eqnarray}
 \left\langle {\Phi _q } \right| = \left\langle {\phi _d } \right|
 \left\langle \vec q  \right|  ~.
\label{e_el.5}
 \end{eqnarray}

Let us consider first the case without 3NF's:
\begin{eqnarray}
&&  \left| U \right\rangle  = tG_0 (1 + P) {\bf O} \left| \Psi_i
\right\rangle
   + tG_0 P\left|  U \right\rangle  \hfill  ~.
\label{e_el.2ab}
\end{eqnarray}

Projecting Eq.~(\ref{e_el.2ab})
on states $|\alpha>$ and $|\beta>$ (in the $|\beta>$ states
 only the screened Coulomb force $V_c^R$ is acting) one gets:
\begin{eqnarray}
\left\langle {{pq\alpha }}
 \mathrel{\left | {\vphantom {{pq\alpha } U}}
 \right. \kern-\nulldelimiterspace}
 {U} \right\rangle  = \left\langle {pq\alpha } \right|t_{N + c}^R
G_0 (1+P){\bf O}\left|
\Psi_i  \right\rangle  + \left\langle {pq\alpha } \right|t_{N + c}^R
G_0 P\left| U
\right\rangle
 \label{e_el.6}
\end{eqnarray}
and
\begin{eqnarray}
&&  \left\langle {{pq\beta }}
 \mathrel{\left | {\vphantom {{pq\beta } U}}
 \right. \kern-\nulldelimiterspace}
 {U} \right\rangle  = \left\langle {pq\beta } \right| t_c^R G_0(1+P)
 {\bf O} \left| \Psi_i
\right\rangle  + \left\langle {pq\beta } \right|t_c^R G_0 P\left| U
\right\rangle  \cr
&&= \left\langle {pq\beta } \right|t_c^R G_0(1+P) {\bf O} \left| \Psi_i
\right\rangle  \hfill
   + \left\langle {pq\beta } \right|t_c^R G_0 P\left|
\alpha'  \right\rangle \left\langle {\alpha' }
 \mathrel{\left | {\vphantom {\alpha'  U}}
 \right. \kern-\nulldelimiterspace}
 {U} \right\rangle  + \left\langle {pq\beta } \right|t_c^R G_0 P\left|
\beta'  \right\rangle \left\langle {\beta' }
 \mathrel{\left | {\vphantom {\beta'  U}}
 \right. \kern-\nulldelimiterspace}
 {U} \right\rangle  \hfill ~.
 \label{e_el.7}
\end{eqnarray}
Since in above equations the photon absorption operator ${\bf {O}}$ 
 comes always with $^3$He bound state therefore 
 $\langle \beta | {\bf O} | \Psi_i \rangle = \langle \beta | P {\bf O} | \Psi_i
\rangle = 0$.  Consequently the first term in
(\ref{e_el.7}) vanishes.
The last term is proportional to
$\left\langle {pq\beta } \right|t_c^R G_0 P\left| \beta  \right\rangle
\left\langle \beta  \right|t_c^R $ and the direct
 calculation of its isospin part gives zero.
 Thus Eq.~(\ref{e_el.7}) reduces to:
\begin{eqnarray}
\left\langle {{pq\beta }}
 \mathrel{\left | {\vphantom {{pq\beta } U}}
 \right. \kern-\nulldelimiterspace}
 {U} \right\rangle  =  \left\langle {pq\beta } \right|t_c^R G_0 P\left|
\alpha'  \right\rangle \left\langle {\alpha' }
 \mathrel{\left | {\vphantom {\alpha'  U}}
 \right. \kern-\nulldelimiterspace}
 {U} \right\rangle ~.
\label{e_el.8}
\end{eqnarray}
Inserting (\ref{e_el.8}) into (\ref{e_el.6})  ones gets:
\begin{eqnarray}
&&  \langle pq\alpha |
 U \rangle  = \langle pq\alpha | t_{N + c}^R G_0(1+P) {\bf O} |
\Psi_i  \rangle  + \langle pq\alpha  | t_{N + c}^R G_0 P |
\alpha'  \rangle \langle \alpha' |
 U \rangle  + \langle pq\alpha  | t_{N + c}^R G_0 P |
\beta'  \rangle \langle \beta' |
 U \rangle   \cr &&
   = \langle pq\alpha  | t_{N + c}^R G_0 (1+P){\bf O} | \Psi_i  \rangle
+ \langle pq\alpha   | t_{N + c}^R G_0 P | \alpha'  \rangle
 \langle \alpha' | U \rangle  \cr &&
   + \langle pq\alpha  | t_{N + c}^R G_0 P | \beta'  \rangle
 \langle \beta'  | t_c^R G_0 P | \alpha'
 \rangle \langle \alpha' | U \rangle ~.
\label{e_el.9}
\end{eqnarray}
Using the completeness relation for the $|\alpha>$ and $|\beta>$ states gives:
\begin{eqnarray}
&&  \left\langle {{pq\alpha }}
 \mathrel{\left | {\vphantom {{pq\alpha } U}}
 \right. \kern-\nulldelimiterspace}
 {U} \right\rangle  = \left\langle {pq\alpha } \right|t_{N + c}^R
G_0(1+P) {\bf O} \left|
\Psi_i  \right\rangle  + \left\langle {pq\alpha } \right|t_{N + c}^R G_0 P\left|
\alpha'  \right\rangle \left\langle {\alpha' }
 \mathrel{\left | {\vphantom {\alpha  U}}
 \right. \kern-\nulldelimiterspace}
 {U} \right\rangle  \hfill \cr &&
   + \left\langle {pq\alpha } \right|t_{N + c}^R G_0 Pt_c^R G_0 P\left| \alpha'
\right\rangle \left\langle {\alpha' }
 \mathrel{\left | {\vphantom {\alpha  U}}
 \right. \kern-\nulldelimiterspace}
 {U} \right\rangle  - \left\langle {pq\alpha } \right|t_{N + c}^R G_0 P\left|
\alpha'  \right\rangle \left\langle \alpha'  \right|t_c^R G_0 P\left|
\alpha''  \right\rangle \left\langle {\alpha'' }
 \mathrel{\left | {\vphantom {\alpha  U}}
 \right. \kern-\nulldelimiterspace}
 {U} \right\rangle  \hfill ~.
  \label{e_el.11}
\end{eqnarray}

When compared with the set resulting from (\ref{e_el.2ab}) for a 
neutron-neutron-proton system there are two new terms in the kernel:
$\left\langle {pq\alpha } \right|t_{N + c}^R G_0 Pt_c^R G_0 P\left| \alpha'
\right\rangle \left\langle {\alpha' }
 \mathrel{\left | {\vphantom {\alpha  U}}
 \right. \kern-\nulldelimiterspace}
 {U} \right\rangle$
 and
 $- \left\langle {pq\alpha } \right|t_{N + c}^R G_0 P\left| \alpha'
\right\rangle \left\langle \alpha'  \right|t_c^R G_0 P\left|
\alpha''  \right\rangle \left\langle {\alpha'' }
 \mathrel{\left | {\vphantom {\alpha  U}}
 \right. \kern-\nulldelimiterspace}
 {U} \right\rangle$. They  are identical to those in (\ref{55}) for the 3N
 continuum and consequently also their evaluation is the same as 
for pd scattering. The vanishing of $|\beta \rangle$-components of the 
 ${\bf O} | \Psi_i \rangle$-state caused, that
instead of three  leading terms as in (\ref{55}), only one leading
term appears, which can be
calculated in a standard way \cite{gol_rep}.

Starting from (\ref{e_el.2a}) and performing analogous steps when the 3NF
is included gives:
\begin{eqnarray}
&&  \left\langle {{pq\alpha }} |
 U \right\rangle  = \left\langle {pq\alpha } | t_{N + c}^R
G_0(1+P) {\bf O} | \Psi_i  \right\rangle
 + \left\langle {pq\alpha } | \frac{1}{2}(P + 1)V_4^{(1)} G_0 (1 + t_{N + c}^R
  G_0 ) (1 + P) {\bf O} | \Psi_i  \right\rangle  \hfill \cr
&&+ \left\langle {pq\alpha } | t_{N + c}^R G_0 P |
\alpha'  \right\rangle \left\langle {\alpha' } |
 U \right\rangle  \hfill \cr &&
   + \left\langle {pq\alpha } | t_{N + c}^R G_0 Pt_c^R G_0 P | \alpha'
\right\rangle \left\langle {\alpha' } |
 U \right\rangle  - \left\langle {pq\alpha } | t_{N + c}^R G_0 P |
\alpha'  \right\rangle \left\langle \alpha'  | t_c^R G_0 P |
\alpha''  \right\rangle \left\langle {\alpha'' } |
 U \right\rangle  \hfill\cr
&& + \left\langle {pq\alpha } | \frac{1}
{2}(P + 1)V_4^{(1)} G_0 (1 + t_{N + c}^R G_0 )P   | \alpha'
\right\rangle \left\langle {\alpha' } | U
\right\rangle  \hfill
 ~.
  \label{e_el.11_3nf}
\end{eqnarray}

Thus adding a 3NF results in one additional leading term,
$\left\langle {pq\alpha } | \frac{1}{2}(P + 1)V_4^{(1)} G_0 (1 + t_{N + c}^R
  G_0 ) (1 + P) {\bf O} | \Psi_i  \right\rangle$, and one additional  
kernel term,
$\left\langle {pq\alpha } | \frac{1}
{2}(P + 1)V_4^{(1)} G_0 (1 + t_{N + c}^R G_0 )P   | \alpha'
\right\rangle \left\langle {\alpha' } | U
\right\rangle$.

The matrix elements  $\left\langle {{pq\alpha }} |  U \right\rangle$
provide transition amplitudes for the two- and three-body breakup of $^3$He.
 Namely, for the two-body breakup of $^3$He the second term in
(\ref{e_el.4}) can be calculated using (\ref{e_el.8}) and
the completeness of the $|\alpha>$ and
$|\beta>$ states, resulting in:
\begin{eqnarray}
&&\left\langle {\Phi _q } \right|P\left|  U \right\rangle =
\langle {\Phi _q } |P | \alpha \rangle \langle \alpha |  U \rangle
+ \langle {\Phi _q } |P | \beta \rangle \langle \beta |  U \rangle \cr
&& = \langle {\Phi _q } |P | \alpha \rangle \langle \alpha |  U
\rangle
+ \langle {\Phi _q } |P t_c^R G_0 P | \alpha' \rangle \langle \alpha'|U\rangle
- \langle {\Phi _q } |P | \alpha \rangle \langle \alpha |
 t_c^R G_0 P | \alpha' \rangle \langle \alpha'| U \rangle  ~.
\label{e_el.4_sec}
\end{eqnarray}
The first and third terms can be obtained  from
 the $|\alpha>$ partial-wave-projected matrix elements using (B.2) of
Ref.~\cite{elascoul}. The second term must be calculated using directly
the 3-dimensional screened Coulomb t-matrix $t_c^R$ according to (D.9)
of Ref.~\cite{elascoul}.

For the three-body breakup of $^3$He the second term in
(\ref{e_el.3}) can be calculated in a similar way  and is given by:
\begin{eqnarray}
&& \left\langle {\Phi _0 } \right|(1 + tG_0 )P
\left| U \right\rangle =
  \langle {\Phi _0 }  | P  | \alpha \rangle \langle \alpha | U \rangle
+ \langle {\Phi _0 }  | P  | \beta \rangle \langle \beta | t_c^R G_0 P
| \alpha' \rangle \langle \alpha'| U \rangle \cr &&
+\langle {\Phi _0 } | \alpha \rangle \langle \alpha | t_{N+c}^R G_0 P| U\rangle
+\langle {\Phi _0 } | \beta \rangle \langle \beta | t_c^R G_0 P|
U\rangle \cr &&
=  \langle {\Phi _0 }  | P  | \alpha \rangle \langle \alpha | U \rangle
+ \langle {\Phi _0 }  | P  t_c^R G_0 P
| \alpha \rangle \langle \alpha| U \rangle
- \langle {\Phi _0 }  | P  | \alpha \rangle \langle \alpha | t_c^R G_0 P
| \alpha' \rangle \langle \alpha'| U \rangle \cr &&
+ \langle {\Phi _0 } | \alpha \rangle \langle \alpha | t_{N+c}^R G_0 P
| \alpha' \rangle \langle \alpha'| U\rangle
+ \langle {\Phi _0 } | \alpha \rangle \langle \alpha | t_{N+c}^R G_0 P
t_c^R G_0 P | \alpha' \rangle \langle \alpha'| U\rangle \cr &&
- \langle {\Phi _0 } | \alpha \rangle \langle \alpha | t_{N+c}^R G_0 P
| \alpha' \rangle \langle \alpha'|
t_c^R G_0 P | \alpha'' \rangle \langle \alpha''| U\rangle \cr &&
+ \langle {\Phi _0 } | t_c^R G_0 P | \alpha \rangle \langle \alpha| U\rangle
- \langle {\Phi _0 } | \alpha \rangle \langle \alpha |
t_c^R G_0 P | \alpha' \rangle \langle \alpha'| U\rangle ~.
\label{e_el.3_sec}
\end{eqnarray}
Here again, the second, fifth and seventh term
 must be calculated using directly
the 3-dimensional screened Coulomb t-matrix $t_c^R$. For the second,
$\langle {\Phi _0 }  | P  t_c^R G_0 P|\alpha \rangle \langle \alpha| U
\rangle$,
and seventh,
$\langle {\Phi _0 } | t_c^R G_0 P | \alpha \rangle \langle \alpha|
U\rangle$, term the calculation follows expressions (D.6), (D.7) and
(D.8) of Ref.~\cite{elascoul}. For the fifth matrix element,
 $\langle {\Phi _0 } | \alpha \rangle \langle \alpha | t_{N+c}^R G_0 P
t_c^R G_0 P | \alpha' \rangle \langle \alpha'| U\rangle$, the
corresponding expressions of
Ref.~\cite{elascoul} are (A.19) and (B.1).
The calculation of the remaining, $|\alpha>$ partial-wave-projected 
matrix elements
 in (\ref{e_el.3_sec}) follows (B.1) of
  Ref.~\cite{elascoul}.

\section{Summary}
\label{sumary}

We extended our  approach to include the pp Coulomb
force into the momentum space 3N Faddeev calculations, presented
in Ref.~\cite{elascoul,brcoul} for elastic pd scattering and  breakup
in case when only pairwise forces are acting,
to include also a 3NF and to treat reactions induced by interaction of
electromagnetic probes with the $^3$He nucleus.
It is based on a standard
 formulation for short range forces and relies on the screening  of the
long-range Coulomb interaction. In order to avoid  all uncertainties
connected with the application of the partial wave expansion,
unsuitable when working
with long-range forces, we apply directly the 3-dimensional pp screened
Coulomb t-matrix.

For each reaction considered in the present study:
elastic pd scattering and breakup, two-
and three-body decay of the $^3$He nucleus induced by real or virtual
photons, the resulting coupled set of integral
equations in the finite space of  $|\alpha>$ channels only, incorporates
the contributions of the pp Coulomb interaction from all partial wave
states up to infinity. Adding a  3NF results in a set of Faddeev-type
equations with the same
structure as in the case when only 2N interactions and pp Coulomb
force are acting. On top of that for each reaction one new
contribution  in the
leading term and in the kernel appears. These two additional terms
have the same form independent if the pp Coulomb force is acting or not.

Solutions of the resulting Faddeev equations in form of partial-wave
projected matrix elements,  together with the additional matrix
elements calculated directly with the 3-dimensional screened Coulomb
t-matrix, provide  transiton amplitudes from which numerous
observables can be calculated.

Since in \cite{elascoul,brcoul} the practical feasibility of our
formulation has been documented  in case of pd elastic scattering and
breakup, the presented extension of similar structure  
will also  be feasible and will allow  to
apply that approach  with the complete nuclear Hamiltonian to analyses 
of numerous  data from 3N hadronic and electromagnetic reactions.

\section*{Acknowledgments}
This work was supported by the Polish 2008-2011 science funds as a
 research project No. N N202 077435.
 It was also partially supported by the Helmholtz
Association through funds provided to the virtual institute ``Spin
and strong QCD''(VH-VI-231).

\end{document}